\documentclass[prl,twocolumn,showpacs,preprintnumbers,amsmath,amssymb]{revtex4}
\usepackage{graphicx}
\usepackage{dcolumn}
\usepackage{bm}

\begin{document}


\title{Fine-tuning the functional properties of carbon nanotubes via the interconversion of encapsulated molecules}

\author{H. Shiozawa*$^1$, T. Pichler$^1$, C. Kramberger$^1$, A. Gr\"uneis$^1$, M. Knupfer$^1$, B. B\"uchner$^1$, V. Z\'olyomi$^{2,3}$, J. Koltai$^{3}$, J. K\"urti$^3$, D. Batchelor$^4$, and H. Kataura$^5$}
\date{\today}
\address{$^1$ IFW Dresden, P.O. Box 270116, D-01171 Dresden, Germany
\\ $^2$ Research Institute for Solid State Physics and Optics of the
Hungarian Academy of Sciences, P. O. B. 49, H-1525, Budapest,
Hungary
\\ $^3$ Department of Biological Physics, E\"otv\"os University, P\'azm\'any
P\'eter s\'et\'any 1/A, H-1117 Budapest, Hungary
\\ $^4$ Universit\"at W\"urzburg, BESSY I\hspace{-.1em}I, D-12489 Berlin, Germany
\\ $^5$ Nanotechnology Research Institute, AIST,
Tsukuba 305-8562, Japan}

\begin{abstract}
Tweaking the properties of carbon nanotubes is a prerequisite for
their practical applications. Here we demonstrate fine-tuning the
electronic properties of single-wall carbon nanotubes via filling
with ferrocene molecules. The evolution of the bonding and charge
transfer within the tube is demonstrated via chemical reaction of
the ferrocene filler ending up as secondary inner tube. The charge
transfer nature is interpreted well within density functional
theory. This work gives the first direct observation of a
fine-tuned continuous amphoteric doping of single-wall carbon
nanotubes.
\end{abstract}


\maketitle

Filled single-wall carbon nanotubes (SWCNT), so called peapods,
are of great research interest. Their unique nanoscale structures
have a potential to bring forth novel electronic and magnetic
properties. After the pioneering works of filling with C$_{60}$
and other fullerenes or fullerene derivatives \cite{Smith1998,
Bandow2001, Shiozawa2006C60, Shiozawa2006Dy}, organic molecules
are becoming promising candidates as a functional filler
\cite{Takenobu2003, Fujita2005, Yanagi2005}. Since their $\pi$
conjugated orbital states are highly sensitive to the chemical
environments, the incorporated organic molecules can react with
the surrounding, modifying the properties of the SWCNT.
Specifically, organometallics are in the focus of attention as a
filling material due to their large variation of the chemical
properties depending upon the metal atom species. Recent
theoretical studies on metallocene filled SWCNT highlighted that
the variety of the metal center in between two planar aromatic
ligands yields tunable carrier doping of SWCNT \cite{Lu2004,
GarciaSuarez2006, Sceats2006}. Experimentally, the filling of
SWCNT with the metallocenes has been so far observed with
high-resolution transmission electron microscopy (HRTEM) with
cobaltocene (CoCp$_2$) \cite{Li2005} and ferrocene (FeCp$_2$)
\cite{Guan2005, Li2006}. Tube-diameter selective filling and
resulting energy shifts of the corresponding fluorescence peaks
were observed for the cobaltocene peapods (CoCp$_2$@NT). The HRTEM
studies on a FeCp$_2$@NT reported the conversion of the
FeCp$_2$@NT to double-wall carbon nanotubes (DWCNT)
\cite{Guan2005} or to iron clusters \cite{Li2006,LiNano2006}.
Although these experimental works using tube materials synthesized
by arc-discharge \cite{Guan2005, Li2006, LiNano2006} or
high-pressure CO conversion (HiPCO) methods \cite{Li2005} allowed
ones to probe the local properties of a certain-type of individual
filled tubes, it has been highly demanding to obtain a bulk scale
filling of SWCNTs with a defined diameter distribution, in order
to get further insight into the bulk electronic and chemical
properties and to checkup the theoretical predictions
\cite{Lu2004, GarciaSuarez2006, Sceats2006}. Very recently, the
bulk-scale encapsulation of FeCp$_2$ inside SWCNTs was
accomplished and an extended use of filled nanotubes in
nanochemistry was proposed: the reaction process of FeCp$_2$
filler precursors towards secondary inner tube growth was
identified and controlled on a bulk scale \cite{Shiozawa}.

In the present work, we evidence that the reaction process of the
filler molecule can be utilized for tuning the properties of
SWCNT. The evolution of the chemical composition of the filler
during the conversion to DWCNT is traced by observing the iron
core-level absorption. The Fe 2$p$ absorption spectra show a clear
sign of encapsulated FeCp$_2$ and their decomposition by vacuum
annealing, which is complementary to the previous photoemission
and HRTEM study \cite{Shiozawa}. The C 1$s$ core photoemission and
absorption study unravels interactions between nanotube carbon
atoms and filler molecules in the FeCp$_2$@NT, which diminish with
the decomposition of FeCp$_2$. Further, a fully continuous
amphoteric doping of SWCNT via chemical reactions of the filler is
evidenced by tracing Van Hove singularity (VHS) peaks in
valence-band photoemission. Both n-type and p-type doping of the
outer tube are observed as bipolar energy shifts of the VHS peaks
relative to those for the pristine SWCNT. A comprehensive analysis
of the charge distribution within the FeCp$_2$@NT and the empty
DWCNT gives very good agreement between experiment and theory.
This work unveils the bonding and charge transfer type
interactions between carbon nanotubes and filler molecules and
pioneers the use of chemical reactions of encapsulated molecules
in tuning the functional properties of SWCNT. This paves the way
towards nanoscale electronic devices based on endohedrally
functionalized carbon nanotubes.

The FeCp$_2$@NT films with a bulk filling factor of about 35 \%
were prepared according to the previous work \cite{Shiozawa},
using the laser ablation SWCNT material with tube diameters of 1.4
nm $\pm$ 0.1 nm \cite{Kataura2002}. A conversion of the
FeCp$_2$@NT to the iron doped DWCNT was carried out via annealing
at 600 $^\circ$C in vacuum. In a similar way the same batch of the
FeCp$_2$ was transformed into empty DWCNT by annealing at 1150
$^\circ$C for 1 hour. The ultra-violet photoemission spectroscopy
was performed with monochromatic He I radiation (21.22 eV) using a
hemispherical SCIENTA SES 200 photoelectron energy analyzer at the
IFW Dresden. The experimental resolution (better than 20 meV) and
the Fermi energy were determined from the Fermi edge of clean
Molybdenum substrates. The x-ray absorption experiment was carried
out at beamline UE 52 PGM, Bessy I\hspace{-.1em}I. Either drain
current or partial electron yield was collected with an overall
resolving power better than 10000. The base pressure in the setup
was kept below 5 $\times$ 10$^{-10}$ mbar. Calculations were
performed using density functional theory in the local density
approximation with the VASP package \cite{RefVasp}. The cutoff
energy of the plane-wave basis set was set to 350 eV. Geometry
optimizations were performed until the forces on all atoms were
below 0.003 eV/\AA. Atomic Bader charges were evaluated using an
external utility \cite{RefHenkelmanCMS}.


\begin{figure}
\includegraphics[width=65mm]{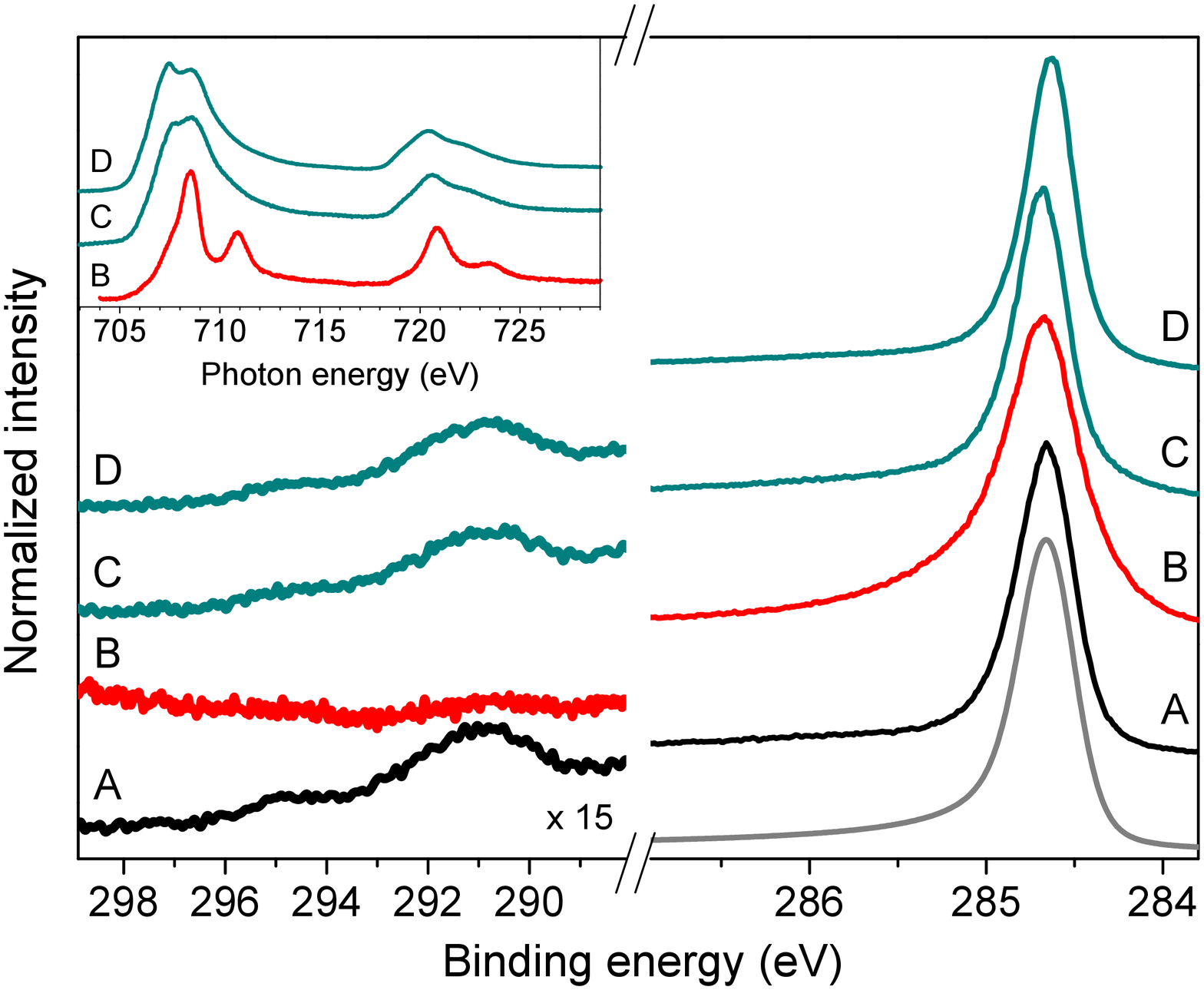}
\caption{C 1$s$ photoemission spectra for samples, A. SWCNT, B.
FeCp$_2$@NT, C. Fe doped DWCNT (600$^\circ$C for 2 hours) and D.
Fe doped DWCNT (600$^\circ$C for 22 hours). Gray line is a Doniach
Sunjic profile convoluted with Gaussian, reproducing well the C
1$s$ peak of the SWCNT. Inset: Fe 2$p$ absorption spectrum for the
samples B, C and D.}
\end{figure}

\begin{figure}
\includegraphics[width=65mm]{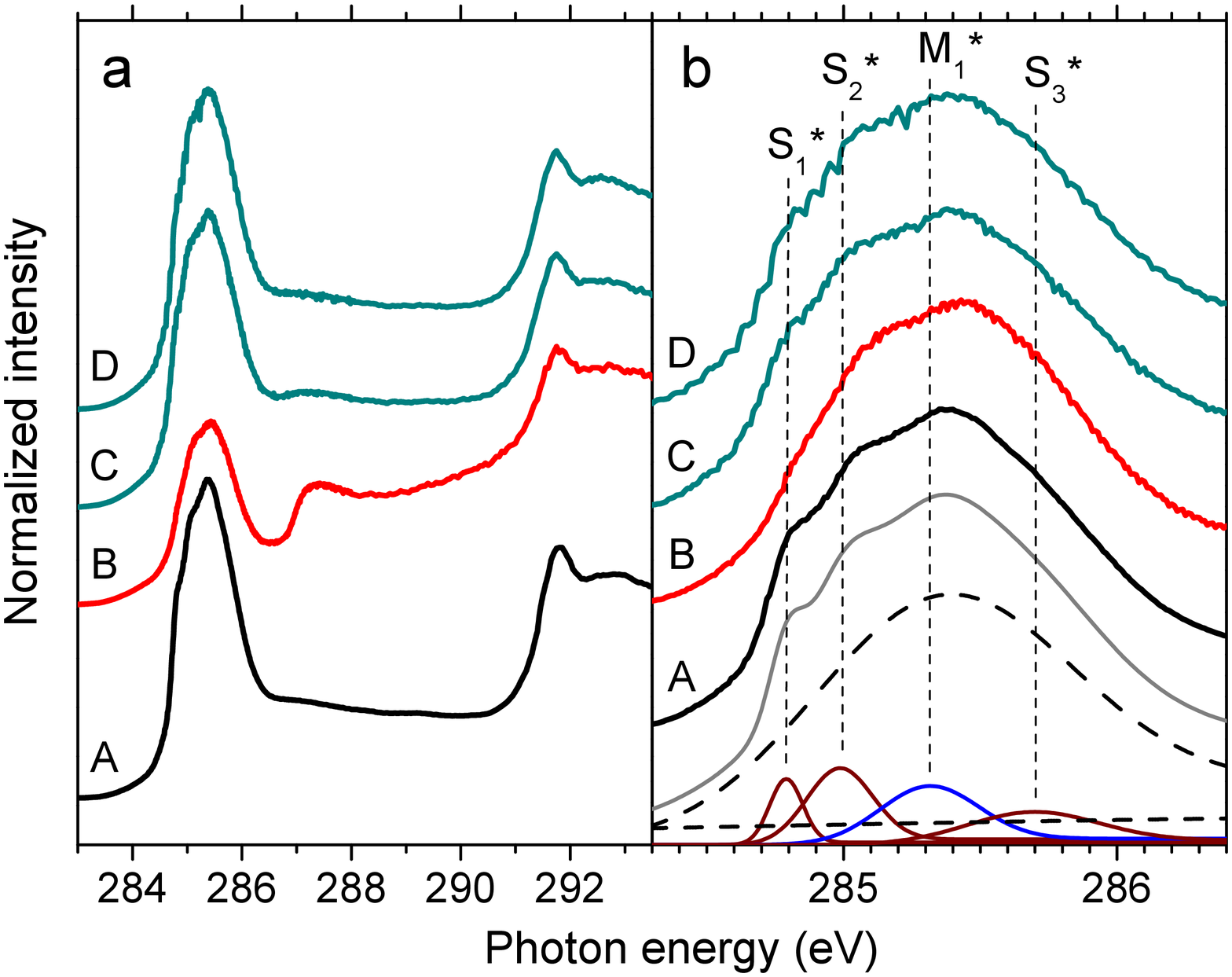}
\caption{a) C 1$s$ absorption spectra for the samples A. SWCNT, B.
FeCp$_2$@NT, C. Fe doped DWCNT (600$^\circ$C for 2 hours) and D.
Fe doped DWCNT (600$^\circ$C for 22 hours). b) Van Hove
singularity peaks at $\pi$ resonance edge on an expanded scale.
Gray line shows the result of line shape analysis for the SWCNT
(Ref. \onlinecite{Kramberger}).}
\end{figure}

The Fe 2$p$ absorption traces the chemical environment of the iron
atoms. The inset of Fig. 1 compares the Fe 2$p$ absorption spectra
between the pristine FeCp$_2$@NT sample and the iron doped DWCNT
samples. For the FeCp$_2$@NT two Fe 2$p$ spin-orbit doublets are
observed. This is very similar to the spectrum for pristine
FeCp$_2$ observed previously \cite{Hitchcock1990} and evidences
for the FeCp$_2$ molecules encapsulated without decompositions or
significant modifications in structure. The iron doped DWCNT
samples in turn show a set of broad and asymmetric doublets
adhering to each other corresponding to variant chemical
compositions of iron as observed previously with Fe 2$p$
photoemission \cite{Shiozawa}. Such a drastic change in the Fe
2$p$ absorption is a direct observation of decomposition of the
FeCp$_2$ molecules.

Further, the C 1$s$ photoemission and absorption uncovers the
nature of nanotube - FeCp$_2$ interactions. The main panel of Fig
1 shows the C 1$s$ photoemission spectra for the SWCNT (spectrum
A), FeCp$_2$@NT (spectrum B) and Fe doped DWCNT samples (spectrum
C and D). The C 1$s$ main peak of the FeCp$_2$@NT is obviously
broader and more asymmetric than that of the SWCNT. This feature
is better recognized by fitting them to Doniach Sunjic profile.
The derived width and asymmetric parameters are 0.175 and 0.147,
respectively, which are factor 1.83 and 1.38 larger than those for
the SWCNT. In turn, those for the Fe doped DWCNT samples almost
recover the values for the SWCNT. Interestingly, this anomaly of
FeCp$_2$@NT is concomitant to a decrease of the $\pi$ plasmon peak
located around 291 eV binding energy in the shake up region,
possibly indicating reduced C $\pi$ components. This is further
investigated with the C 1$s$ absorption plotted in Fig. 2a. The C
1$s$ absorption spectrum for the pristine SWCNT exhibits $\pi$
dominant peak located at photon energy around 285.4 eV and
$\sigma$ dominant structures above 291.8 eV. In the $\pi$ peak
plotted on the expanded scale in Fig 2b, the fine structures
corresponding to the VHS are observed for all the samples. Those
are reproduced well by the line shape analysis according to the
previous work \cite{Kramberger}, as depicted at the bottom in Fig
2b. This observation confirms high purity of the samples. The C
$sp^2$ feature consisting of the $\pi$ and $\sigma$ structures is
drastically modified in the FeCp$_2$@NT. The $\pi$ peak intensity
relative to the $\sigma$ structures is strongly reduced. Instead,
a new peak appears at 287 eV. This originates from the FeCp$_2$
molecule which has a main peak structure at 287 eV corresponding
to C-H bonds \cite{Ruhl1988}. It should be noted, however, that
there must be much less pronounced features of the raw FeCp$_2$
observed accounting for the estimated concentration of the
FeCp$_2$ consisting carbons \cite{Shiozawa}. In addition to the
reduced $\pi$ peak intensity, therefore, this can be attributed to
the nanotube carbon valence states modified by FeCp$_2$ adsorption
on the inner wall so as to project the spectral weight of the
filler molecule on the C 1$s$ absorption. The diminished fraction
of the $\pi$ peak is transferred to this additional peak. This
also indicates stronger interaction between FeCp$_2$ $\pi$
conjugated orbital and nanotube $\pi$ orbital, which is
rationalized in terms of the geometry of the $\pi$ and $\sigma$
orbital within the tube. Orbital mixing of the nanotube $\pi$ and
FeCp$_2$ $\pi$ conjugated orbital must be responsible for this
feature. This can cause a significant modification of the nanotube
valence states. The features are very similar to those observed
previously for hydrogenated or oxidized SWCNT \cite{Nikitin2005}.
In general, the additional bonding to nanotube wall yields more
$sp^3$ like carbon atoms, which results in reducing the $\pi$
dominant signal \cite{Stohr}. This also explains the reduced $\pi$
plasmon shake up in the C 1$s$ photoemission of the FeCp$_2$@NT.
Additionally, those $sp^3$ like features observed in both
photoemission and absorption diminish rapidly after 2 hours
annealing, indicating the detachment of those bonding as a
consequence of the decomposition of the FeCp$_2$ molecule as
observed with the Fe 2$p$ absorption.

\begin{figure}
\includegraphics[width=70mm]{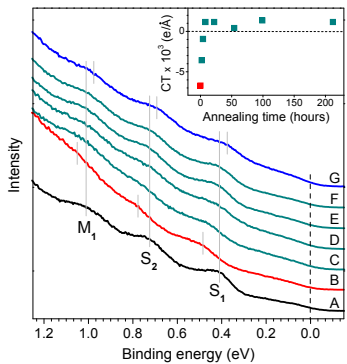}
\caption{He I photoemission spectra for A. pristine SWCNT, B.
FeCp$_2$@NT, C-F. Fe doped DWCNT or FeCp$_2$@NT annealed at
600$^\circ$C for 2, 8, 54, 212 hours, respectively, G. FeCp$_2$@NT
annealed at 1150$^\circ$C for 1 hour. Inset: annealing time
dependence of charge transfer density (CT) for the outer tube in
units of e/$\rm \AA$.}
\end{figure}

We now turn to investigate the degree of charge transfer between
the tube and the filler material. This is done by observing the
VHS peaks in photoemission. The valence band photoemission
spectrum of the pristine SWCNT exhibits three peaks corresponding
to the VHS of semiconducting tubes ($S_{\rm 1}$, $S_{\rm 2}$) and
metallic tubes ($M_{\rm 1}$) \cite{Ishii2003}, respectively, as
shown in Fig. 3. For the FeCp$_2$@NT those peaks are significantly
shifted towards higher binding energies. From a close similarity
to alkali metals doped SWCNT materials \cite{Rauf2004,
Larciprete2005}, this behavior is safely attributed to the
electron charge transfer from the FeCp$_2$@NT to SWCNT. This was
reported in the previous work \cite{Shiozawa} where a charge
transfer value for the FeCp$_2$@NT was derived from the filling
factor and energy shifts of the VHS peaks by comparing them with
the ones for potassium doped SWCNT. Upon annealing at 600
$^\circ$C, the VHS peaks are gradually shifted towards lower
binding energies and become stable after 22 hours. Surprisingly,
the reached energy positions are lower than those for the pristine
SWCNT. At this stage the material is transformed into the iron
doped DWCNT as reported previously \cite{Shiozawa, footnote1}. In
turn, upon 1150 $^\circ$C annealing, we get the empty DWCNT,
without any iron atoms, whose VHS peaks are further shifted
towards lower energies. This yields a significant conclusion that
there is electron transfer from the outer tube to the inner tube
in the empty DWCNT. The direction of the charge transfer is the
same as reported in a recent theoretical work \cite{pssbCTpaper,
Zolyomi_condmat}. Moreover, the difference between the empty and
iron doped DWCNT highlights electron charge transfer from the iron
materials to the outer tubes. This electron doping should be both
to the inner and outer tubes and the averaged doping level
observed with VHS peak shifts depends on the chemical composition
of fillers.

Finally, a quantitative analysis of the charge transfer between
the filler and the SWCNT is given both experimentally and
theoretically. According to the previous work \cite{Shiozawa} the
number of electrons transferred from the fillers to the SWCNT was
evaluated from the VHS peak shifts. The inset of Fig. 3 shows the
charge transfer density (CT) of the outer tube along the tube axis
for FeCp$_2$@NT samples annealed at 600 $^\circ$C for different
durations. The CT has a maximum minus value of -0.0067 $\rm e/\AA$
for the pristine sample and drops rapidly at the first few
annealing steps and becomes plus constant around +0.0010 $\rm
e/\AA$ after 8 hours annealing. In contrast, the iron
concentration is much less dependent of annealing and shows only a
factor 0.7 drop in the end. These features suggest that the
chemical status of the filler material is certainly responsible
for the evolution of the CT. For the representative two filled
samples, FeCp$_2$@NT and the empty DWCNT, the possible filled
structures are available \cite{Shiozawa}, thus, the CT for the
actual samples can be extrapolated to those for 100 \% filled
materials. The results are $-0.0180$ $\rm e/\AA$ for FeCp$_2$@NT
and $+0.0286$ $\rm e/\AA$ for DWCNT.

On the theoretical side, we have calculated the CT between
ferrocene and a host (9,9) nanotube. This tube has a diameter
which is a good representation of our experimental sample and is
highly suited for optimal ferrocene filling. We have examined both
the case of the standing and the lying FeCp$_{2}$, assuming van
der Waals distance between the FeCp$_{2}$ and the nanotube wall.
The geometries of the nanotube and the FeCp$_{2}$ were optimized
separately, and the optimal distance between them was obtained by
finding the minimum of the total energy as a function of the
distance. A full geometry optimization was performed subsequently,
but no essential change was found in the CT values. Both ferrocene
alignments yielded a transfer of electrons from the FeCp$_{2}$ to
the nanotube, in full agreement with the measurements. For the
standing and lying geometry, the CT was found to be $-0.0020$ $\rm
e/\AA$ and $-0.0029$ $\rm e/\AA$, respectively, far less than the
experimental $-0.0180$ $\rm e/\AA$. In comparison, we have also
calculated the CT for the case of the (4,4)@(9,9) DWCNT, which
yielded $+0.0065$ $\rm e/\AA$ for the outer wall. Once again, the
direction of the charge transfer is in full agreement with the
experiment, and in this case the magnitude is also closer. It
should be noted that a previous study \cite{pssbCTpaper,
Zolyomi_condmat} has revealed that the inter-molecular H\"uckel
model \cite{IMH} yields about a factor 2 larger magnitude for the
CT in double walled tubes than the LDA does; in the case of the
(4,4)@(9,9) the CT was found to be $+0.0138$ $\rm e/\AA$ with this
model. The latter value is much closer to the experimental value
of $+0.0286$ $\rm e/\AA$.

In the case of the ferrocene filled tubes, even applying a factor
2 correction yields far too low CT. This discrepancy might suggest
that the FeCp$_{2}$ effectively binds into the tube wall: the
theoretically considered scenario -- i.e. van der Waals distance
between the FeCp$_2$ and the nanotube wall -- allows only for weak
inter-molecular interactions leading to small CT, while a strong
-- possibly covalent -- bonding may allow for a larger CT between
the two subsystems. We have performed further calculations to
examine this possibility by studying a simple test geometry in the
parallel orientation where the 2 hydrogens nearest to the tube
wall are removed from the FeCp$_{2}$ which then binds into the
wall by the thus appearing dangling bonds \cite{footnote2}. We
found that in this geometry the CT is indeed much greater, and the
carbon atoms of the nanotube which are involved in the bonding are
of largely $sp^3$ character. The latter result also agrees with
our aforementioned C 1$s$ core-level measurements which showed
that some of the carbon atoms in the tube wall have a strong
$sp^3$ character, indicative of orbital hybridization between the
subsystems. Thus, both (i) the discrepancy between the measured
and calculated CT and (ii) the $sp^3$ character of carbon atoms
found by C 1$s$ spectroscopy, suggest in unison that there should
be some kind of relatively strong, possibly covalent bonding
between the FeCp$_{2}$ and the host nanotube. This is essentially
an indication of possible endohedral functionalization of the
nanotube, which has been suggested to be theoretically possible
recently \cite{YumuraKerteszCM2007}.

In summary, we have evidenced the potential of filling and
subsequent chemical reactions to tune the properties of SWCNT.
Both n-type and p-type doping of the SWCNT were achieved from
different chemical states of the filler. Good agreement was found
between the experimental and theoretical CT values. Our results
show that covalent and noncovalent interactions are responsible
for the endohedral functionalization of the SWCNT.

\begin{acknowledgments}
This work was supported by the DFG Pl 440/3/4, by OTKA grants no.
K60576 and F68852, Hungary, by a Grant-in-Aid for Scientific
Research A (No.18201017) by the Ministry of Education, Culture,
Sports, Science and Technology of Japan. H. S. acknowledges the
Alexander von Humboldt Foundation. C. K. acknowledges the IMPRS.
A. G. acknowledges an individual Marie Curie fellowship. V. Z.
acknowledges the Bolyai fellowship. We thank R. H\"ubel, S. Leger,
R. Sch\"onfelder and H. Klose for technical assistance.
\end{acknowledgments}


\begin{thebibliography}{}
\bibitem{footnote1} There
should be no discernible features of the inner tubes observed
because of their broad density of states.
\bibitem{footnote2}
This test geometry is energetically unfavorable. In the real
system, instead, we expect that the FeCp$_{2}$ binds into the wall
at defect sites. This will be explored in more detail within a
future work.
\end{thebibliography}
\end{document}